\documentclass[iop,apjl]{emulateapj}

\usepackage{natbib}
\slugcomment{To appear in the Astrophysical Journal Letters}

\shorttitle{Granulation signatures in the spectrum of HD\,122563}
\shortauthors{Ram\'irez et~al.}

\newcommand{\feh}{\mathrm{[Fe/H]}}
\newcommand{\teff}{T_\mathrm{eff}}
\newcommand{\logg}{\log g}

\newcommand{\fei}{Fe\,\textsc{i}}
\newcommand{\feii}{Fe\,\textsc{ii}}
\newcommand{\ms}{m\,s$^{-1}$}

\begin{document}

\title{Granulation signatures in the spectrum of the very metal-poor red giant HD\,122563}

\author{I. Ram\'irez\altaffilmark{1,2}, R. Collet\altaffilmark{2}, D. L. Lambert\altaffilmark{3}, 
        C. Allende Prieto\altaffilmark{4,5}, and M. Asplund\altaffilmark{2}}

\altaffiltext{1}{The Observatories of the Carnegie Institution for Science,
                 813 Santa Barbara Street, Pasadena, CA 91101, USA}
\altaffiltext{2}{Max Planck Institut f\"ur Astrophysik,
                 Postfach 1317, 85741 Garching, Germany}
\altaffiltext{3}{McDonald Observatory and Department of Astronomy,
                 University of Texas, Austin, TX 78712-0259, USA}
\altaffiltext{4}{Instituto de Astrof\'isica de Canarias,
                 38205, La Laguna, Tenerife, Spain}
\altaffiltext{5}{Departamento de Astrof\'{\i}sica, Universidad de La Laguna,
                 38206, La Laguna, Tenerife, Spain}

\begin{abstract}
A very high resolution ($R=\lambda/\Delta\lambda=200,000$), high signal-to-noise ratio ($S/N\simeq340$) blue-green spectrum of the very metal-poor ($\feh\simeq-2.6$) red giant star HD\,122563 has been obtained by us at McDonald Observatory. We measure the asymmetries and core wavelengths of a set of unblended \fei\ lines covering a wide range of line strength. Line bisectors exhibit the characteristic C-shape signature of surface convection (granulation) and they span from about $100$\,\ms\ in the strongest \fei\ features to 800\,\ms\ in the weakest ones. Core wavelength shifts range from about $-100$ to $-900$\,\ms, depending on line strength. In general, larger blueshifts are observed in weaker lines, but there is increasing scatter with increasing residual flux. Assuming local thermodynamic equilibrium (LTE), we synthesize the same set of spectral lines using a state-of-the-art three-dimensional hydrodynamic simulation for a stellar atmosphere of fundamental parameters similar to those of HD\,122563. We find good agreement between model predictions and observations. This allows us to infer an absolute zero-point for the line shifts and radial velocity. Moreover, it indicates that the structure and dynamics of the simulation are realistic, thus providing support to previous claims of large 3D-LTE corrections, based on the hydrodynamic model used here, to elemental abundances and fundamental parameters of very metal-poor red giant stars obtained with standard 1D-LTE spectroscopic analyses.
\end{abstract}


\section{Introduction}

The red giant star HD\,122563 is the brightest very metal-poor ($\feh\lesssim-2.5$) object known ($V=6.2$\,mag). Its extreme metal deficiency was first identified by \cite{wallerstein63} and since then a large number of studies concerning its detailed chemical composition have been published \cite[e.g.,][]{sneden73:hd122563,lambert74,ryan96,cayrel04}. Twenty entries more recent than 1990 are listed for HD\,122563 in the PASTEL database of stellar parameters \citep{soubiran10}, which give: $\teff=4602\pm72$\,K, $\logg=1.25\pm0.22$, and $\feh=-2.64\pm0.12$ (mean and standard deviation).

Most elemental abundance studies use one-dimensional model atmospheres in hydrostatic equilibrium. The atmospheres of cool stars, however, are far from static. Velocity and intensity fluctuations due to surface convection are directly observed in the Sun \citep[e.g.,][]{nesis92,rutten04} and have been detected indirectly using high resolution spectra of a number of other stars \citep[e.g.,][]{allende02,gray05,dravins08,kdwarfs-p1}. In the solar photosphere, bright granules containing hot ascending gas are observed separated by intergranular lanes of cool descending gas. The correlation between temperature and velocity fields, as well as the granule/intergranule contrast in velocity and intensity, shape the spectral features so that the line cores of disk-integrated profiles are shifted with respect to their rest wavelengths and the line profiles are asymmetric \citep[e.g.,][]{dravins81}. The latter are granulation signatures that can be observed both in the solar spectrum and in the spectra of distant stars.

Hydrodynamic simulations provide the most realistic approach to the modeling of surface inhomogeneities \citep[e.g.,][]{nordlund82,steffen91,ludwig09,nordlund09:sun}, although non-hydrodynamic modeling is also possible \citep[e.g.,][]{dravins90}. Three-dimensional model atmospheres for the Sun, Procyon, and K-dwarfs have been tested against high quality spectroscopic data \citep{asplund00:iron_shapes,allende02,kdwarfs-p2}, showing in general good agreement. These tests are crucial because they validate the models and allow us to use them with confidence to estimate the impact of ``3D effects'' on stellar abundance and parameter determinations \citep{asplund05:review}.

The most extreme abundance corrections proposed in connection with granulation effects occur at low metallicities. This is mainly because the difference between the 1D and 3D predictions for the mean temperature of the outer layers of metal-poor stars is very large, on the order of 1000\,K \citep[e.g.,][]{collet06,collet07,gonzalez-hernandez10:3d}. This large temperature difference is very important for the formation of features that are temperature sensitive. For example, \cite{collet09} report 3D corrections of $\sim-1$\,dex (i.e., one order of magnitude) for the abundance of nitrogen and oxygen as determined from molecular features, similar to the corrections required for the iron abundance determined from low excitation potential \fei\ lines. They also estimate a correction of about +0.4\,dex for the iron abundance determined from \feii\ lines.

We can only rely on the corrections suggested by 3D models if the hydrodynamic simulations are realistic. Tests of the hydrodynamic simulations include measurements of the profiles of stellar absorption lines both with respect to shape and absolute wavelength. In this \textit{Letter}, we present new spectroscopic observations for the very metal-poor red giant star HD\,122563 and compare them to state-of-the-art 3D model predictions.

\section{Spectroscopic observations, line data, and measurement of line asymmetries} \label{s:observations}

\setcounter{footnote}{0}

We observed HD\,122563 on June 21--25, 2010 with the Robert~G.~Tull coud\'e spectrograph on the 2.7\,m Harlan~J.~Smith Telescope at McDonald Observatory. Using a narrow slit (0.34\,arcsec), the spectral resolution is $R=200,000$ \citep{tull95}. The wavelength coverage of a single exposure consists of 16 non-overlapping pieces of $\simeq20$\,\AA\ each. To increase wavelength coverage we rotated the spectrograph grating to three different positions (setups). In this way we were able to cover about 70\,\% of the 4425 to 5575\,\AA\ spectral window. To increase the signal-to-noise ratio ($S/N$), multiple 20-minute exposures of each setup (from 8 to 16, depending on the setup) were acquired. The total exposure time for all setups was about 10.5 hours. The $S/N$ of the coadded spectrum increases from about 180 at 4425\,\AA\ to 440 at 5575\,\AA, with a median $S/N=340$. The spectroscopic data were reduced and coadded as in \cite{kdwarfs-p1}. The RMS of the wavelength calibration for each exposure is about $5\times10^{-4}$\,\AA\ ($\simeq30$\,\ms). Continuum normalization was done by fitting polynomials to the upper envelope of the data, first in the direction of dispersion and then perpendicular to it, as in \cite{barklem02}. The barycentric correction was determined with a precision better than 1\,\ms\ using a program by S.~Els (private communication).

A radial velocity correction was applied to each setup by comparing the observed core wavelength shift versus line strength relation with the theoretical one (cf.~Sect.~\ref{s:compare}). This procedure is similar to that described in \cite{gray09} but using a theoretical prediction instead of a solar reference. Basically, the zero point of the observed core wavelength shifts is modified until the average difference with the model, for all lines in a given setup, is exactly zero. The absolute radial velocity we derive by applying this method to the three setups is $V_r=-25.39\pm0.09$~km\,s$^{-1}$ (the error bar corresponds to the 1\,$\sigma$ standard deviation of the three velocities derived).\footnote{This absolute radial velocity is not corrected for the star's gravitational redshift, which is, however, expected to be relatively small \citep[e.g.,][]{dravins99}. Metal-poor red giants have typical masses of $\simeq0.75\,M_\odot$ while we estimate a radius of $\simeq23\,R_\odot$ for HD\,122563 based on its observed \textit{Hipparcos} parallax \citep{vanleeuwen07} and the angular diameter obtained with the infrared flux method \citep{ramirez05a}. These values lead to a gravitational redshift of 633\,\ms~$\times[(M/M_\odot)/(R/R_\odot)]\simeq0.021$\,km\,s$^{-1}$ which is smaller than our uncertainty in $V_r$.} The uncertainty in this velocity is most likely due to instrumental and temperature/pressure instabilities (the R.~G.~Tull spectrograph is not in a vacuum).

We identified 58 unblended \fei\ lines in our HD\,122563 spectrum. For these features we adopted the laboratory wavelengths measured by \cite{nave94}, which have a precision of about 20\,\ms, as estimated from the uncertainty in the wavenumber of the relevant energy levels. The line synthesis presented in Sect.~\ref{s:hydro} was done using transition probabilities measured in the laboratory by a number of groups and homogenized following the recommendations by \cite{lambert96}. The van der Waals damping constants used are those by \cite{barklem00}, except for five lines in which we used the values obtained from the classical Uns\"old formula.

Line bisectors were measured in our HD\,122563 spectrum (and also in an identical fashion in the theoretical line profiles, Sect.~\ref{s:hydro}) by first resampling the spectral lines, using cubic spline interpolations, to iso-flux points on both wings of the lines. The midpoints of horizontal segments across the spectral line wings, which define the line bisector, were then computed. The wavelengths of observed spectral line cores were determined using the 7 data points closest to the flux minimum, which corresponds to a window of about 3.8~km\,s$^{-1}$, and fitting a fourth order polynomial. Core wavelength shifts (hereafter given as a velocity, i.e., $\Delta v_c=c\,\Delta\lambda/\lambda$, where $c$ is the speed of light and $\Delta\lambda=\lambda_\mathrm{obs}-\lambda_\mathrm{lab}$) were then derived using the adopted laboratory wavelengths. Observational errors for the line bisectors were computed following \cite{gray83}.

\begin{figure}
\epsscale{1.2}
\includegraphics[width=9cm,bb=90 360 620 765]{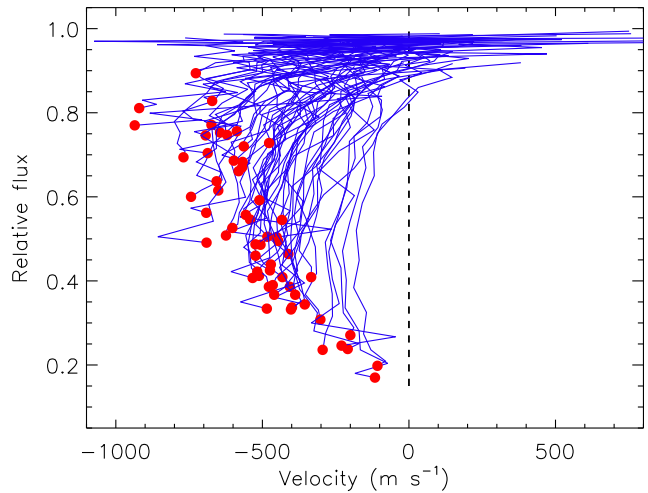}
\caption{Bisectors of 58 \fei\ lines in our spectrum of HD\,122563. Filled circles show the location of the line cores. The dashed line is at zero velocity.}
\label{f:bisectors}
\end{figure}

The observed line bisectors are shown in Fig.~\ref{f:bisectors}, where we emphasize the location of the line cores. The cores of the strongest lines are blueshifted by about $-100$~\ms\ while those of weaker lines show more blueshift, up to $-900$~\ms\ (note that the absolute velocity scale was determined using 3D model predictions, as explained in Sect.~\ref{s:compare}). The overall shape of the bisectors resembles the letter C, although it varies with line strength. Strong (weak) lines resemble more the lower (upper) part of the letter C. This is not very different from the case of cool metal-rich stars, implying that low metallicity does not affect the basic behavior of bisector shapes \citep[see also][]{gray08}. The strongest lines in the spectrum of HD\,122563 (residual flux of about 0.2) are more symmetric and their bisectors have a span of about 100\,\ms. The weaker lines (residual flux of about 0.8), on the other hand, have a bisector span of about 800\,\ms. In a hydrostatic atmosphere, the lines would be symmetric and not shifted. This hypothetical case is illustrated with the dashed line in Fig.~\ref{f:bisectors}. 

\section{Hydrodynamic simulation} \label{s:hydro}

Collet et~al.\ (in preparation; see also \citealt{collet09}) used the \textsc{stagger} code by \cite{nordlund95} to compute a 3D radiative-hydrodynamic model with stellar parameters $T_\mathrm{eff}=4600$~K, $\log{g}=1.6$~[cgs], and scaled solar chemical composition with [Fe/H]$=-3$. The stellar parameters were taken from \cite{barbuy03} with a slight revision of the surface gravity based on the parallax determination of HD122563 from the \textit{Hipparcos} data reduction by \cite{vanleeuwen07}.

The 3D model was computed using a discrete $480{\times}480{\times}240$ grid representing a cube of physical dimensions $3700{\times}3700{\times}1100$~Mm$^3$. The equations of mass, energy, and momentum conservation were solved together with the radiative transfer equations adopting realistic equation of state and opacities. The simulation sequence used for the present work consists of 51 snapshots that cover about 150~hours of stellar time and it typically contains about 10~granules at the surface. The evolution timescale of the granules is about 90 hours. For more details, see \cite{collet09}.

We used the 3D model to synthesize all 58 \fei\ lines shown in Fig~\ref{f:bisectors}. For the line synthesis we adopted a coarser grid of $50{\times}50{\times}125$ points to which we interpolated the original simulation. A number of tests were performed with a higher-resolution grid of $100{\times}100{\times}125$ points to ensure that this had no significant impact on the shapes of the synthetic line profiles.

\section{Line profiles: synthetic vs.\ observed} \label{s:compare}

The synthetic 3D profiles were convolved with a Gaussian instrumental profile of $\mathrm{FWHM}=1.5$~km\,s$^{-1}$, which corresponds to the spectral resolution of our data. From a $\chi^2$ line profile fitting of about 30 \fei\ lines with the highest local $S/N$, with $V\sin i$ and iron abundance as free parameters, we obtained $V\sin i=3.2\pm0.6$~km\,s$^{-1}$. For the $\chi^2$ fitting we convolved the disk-integrated profiles \citep[e.g.,][p.\,370]{gray92:book}. To calculate the disk-integrated profiles for comparison with the observational data we applied a $V\sin i=3.2$~km\,s$^{-1}$ to the emergent intensities. A detailed abundance analysis of our HD\,122563 spectrum using the 3D model will be provided in a later publication.\footnote{According to the preliminary analysis by \cite{collet09}, the 3D LTE iron abundances determined from \fei\ lines show a trend with excitation potential, going from about $\log\epsilon\,\mathrm{(Fe)}=4.0$ at $\chi=0$\,eV to $\log\epsilon\,\mathrm{(Fe)}=4.6$ at $\chi=4$\,eV, with some line-to-line scatter comparable to the one from classical 1D analyses.}

\begin{figure}
\epsscale{1.2}
\includegraphics[width=9.1cm,bb=90 370 480 572]{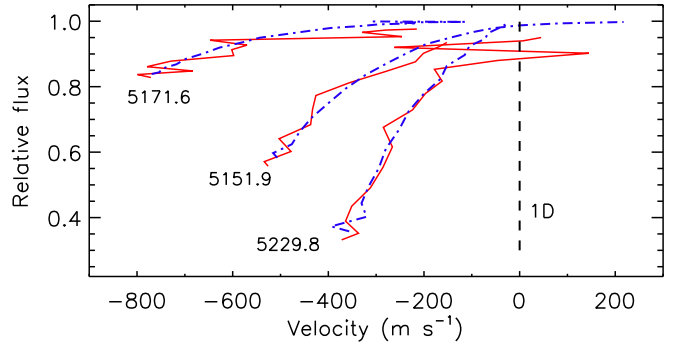}
\caption{Bisectors of three representative \fei\ lines in the spectrum of HD\,122563 (solid lines). Error bars ($\simeq30-150$~\ms\ from core to continuum) have been omitted for clarity. The dot-dashed lines correspond to 3D model predictions. The observed lines in this figure have been shifted to match the theoretical line cores and thus allow a better comparison of bisector shapes. The dashed line at zero velocity represents 1D model predictions.}
\label{f:bis}
\end{figure}

In Fig.~\ref{f:bis} we show a comparison of observations and model for 3 representative \fei\ lines. The observed lines have been shifted by about +30, +30, and --100\,\ms\ (from strongest to weakest line) to make their core velocities agree with the theoretical values. This allows a better comparison of line bisector \textit{shapes} because it removes the uncertainties in the measured core wavelengths and rest laboratory wavelengths. We discuss the core wavelength shifts below and show that the artificial shifts applied here are within the observational errors.\footnote{We avoid the comparison of line bisectors averaged in groups of line strength because this leads to a certain loss of information. The detailed shapes of lines of similar strength can be different due to differences in the properties of the atomic transition involved. Averaging is justified in cases where the $S/N$ is low, the spectral lines are severely blended, or the granulation signatures are too weak \cite[e.g.,][]{kdwarfs-p1}.} The average difference in bisector velocities (model minus observations for all 58 \fei\ lines) is $10\pm71$\,\ms\ (hereafter, error bars correspond to the 1\,$\sigma$ scatter). This number, however, depends on the relative flux. Near the continuum, the bisector point-by-point difference is about $-30\pm145$~\ms\ while it is $9\pm33$~\ms\ at a residual flux of about 0.2. At all flux values the difference is consistent with zero within the $1\,\sigma$ scatter. The formal error in the observed bisector velocities shows a residual flux dependency remarkably similar to that described above; i.e., it ranges from about 30 to 150~\ms\ from core to continuum. Thus, the scatter seen in the differences between theoretical and observed bisectors is fully compatible with the $1\,\sigma$ observational errors.

\begin{figure}
\epsscale{1.2}
\includegraphics[width=8.6cm,bb=60 365 480 930]{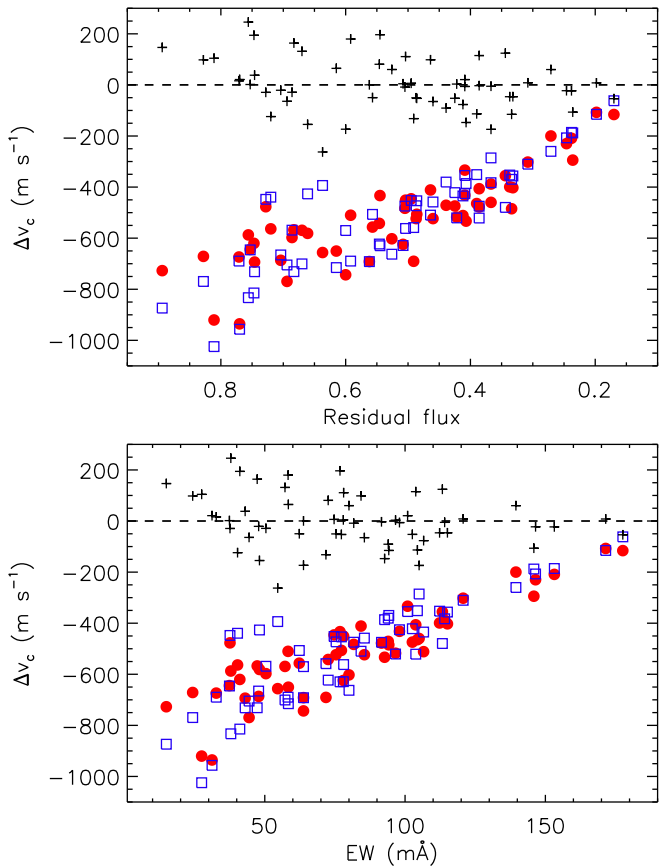}
\caption{Core wavelength shifts of 58 \fei\ lines in the spectrum of HD\,122563 (filled circles) and those predicted by the 3D model (open squares) as a function of residual flux (top panel) and equivalent width (bottom panel). The residuals (observation minus model) are shown with crosses. The dashed line is at zero velocity.}
\label{f:shifts}
\end{figure}

Measurements of core wavelength shifts ($\Delta v_c$) are in principle more precise than line bisectors, since only a small portion of the line profile is examined, thus lowering the probability of being affected by blends. In Fig.~\ref{f:shifts} we show the relation between $\Delta v_c$ and line strength. The so-called ``third signature'' of stellar granulation \citep{gray09}, i.e., the increased blueshift with decreasing line strength, is clearly detected in the data and very closely reproduced by the 3D model. The third signature is due to the fact that weaker lines form in deeper layers, where the granulation velocities and intensity contrast are larger. The line-to-line scatter seen in Fig.~\ref{f:shifts} increases towards weaker lines, an effect that is present also in the 3D model predictions. It is tempting to conclude that this increased scatter is real, but given the timescale of the evolution of granules in comparison with the duration of the simulation and the possibility that line cores of weaker lines are more difficult to measure (but see below), we must refrain from making a strong statement about this finding. Note also that the variation of the continuous opacity over the wavelength range covered by our data is substantial, which introduces wavelength dependent shifts that might contribute to the observed scatter \citep[e.g.,][]{dravins81}. The difference between observations and model predictions for $\Delta v_c$ has a 1\,$\sigma$ scatter of 104\,\ms.

To estimate the error in the measured $\Delta v_c$ values we artificially introduced Gaussian noise to the theoretical profiles to simulate the local $S/N$ of the data. We took into account the variation of the $S/N$ across the line profile since it varies according to the amount of flux detected at each wavelength and it is therefore lower at the line core compared to the local continuum. This experiment was performed after degrading the theoretical profiles to the spectral resolution of our data and convolving it with a rotational profile with the $V\sin i$ we derived for HD\,122563. Finally, for each line tested we added in quadrature the error introduced by the uncertainty in the measured laboratory wavelengths (20\,\ms) and the estimated error of the wavelength calibration of the data (30\,\ms). The $\Delta v_c$ values obtained from 1,000 of these tests, averaged over the 58 \fei\ lines, have a $1\,\sigma$ scatter of $\simeq108$~\ms, nearly independent of line strength but mildly dependent on wavelength, reflecting the variation of the $S/N$ across our observed spectrum. This estimated observational error is very similar to the scatter seen in the comparison of model predictions and observations.

\section{Summary and discussion}

We have observed the very metal-poor red giant star HD\,122563, a primer for studies of the Galactic stellar halo, at very high resolution and high signal-to-noise ratio. The high quality of our data allowed us to measure the asymmetries and core wavelength shifts of a number of \fei\ lines, which clearly revealed the signatures of granulation. A state-of-the-art 3D LTE radiative-hydrodynamic simulation was used to synthesize the same set of lines. Using line profile fitting we derived a projected rotational velocity $V\sin i=3.2\pm0.6$~km\,s$^{-1}$ while the analysis of core wavelength shifts allowed us to infer an absolute radial velocity $V_r=-25.39\pm0.09$~km\,s$^{-1}$. The line bisectors and core wavelength shifts computed from the 3D model predictions agree with the observations of HD\,122563 remarkably well. The average difference between observations and model is consistent with zero within the estimated $1\,\sigma$ observational errors. This suggests that the 3D model contains realistic information regarding the atmospheric structure and velocity fields of HD\,122563.

Abundance analyses of HD\,122563 using our 3D model show that standard 1D estimates of elemental abundances can be off by up to one order of magnitude \citep{collet07,collet09}. Qualitatively similar results have been obtained with a simulation computed by an independent group \citep{ivanauskas10,dobrovolskas10}, although their abundance corrections tend to be smaller. The large 3D corrections will have an impact on our interpretations of stellar abundance data for halo stars and may lead us to revise their connection with early enrichment of the Galaxy, supernovae yields, and internal mixing processes. The observations presented in this \textit{Letter} and their comparison to the 3D model predictions suggest that such large 3D effects are reliable and must therefore be seriously considered in all studies of metal-poor stellar abundances.

Admittedly, the largest 3D corrections apply to abundances derived from molecular features whereas we have analyzed only \fei\ lines. The \fei\ lines provide one sampling of the atmospheric structure with a particular weighting over the different structures and velocity fields. Other species, particularly the molecules, will provide different weighting and should in principle be investigated before claiming that the 3D models are adequately justified for all abundance analyses. Unfortunately, in our spectral range the molecular lines are too weak for this type of work. We note, however, that our \fei\ lines form over a wide range of depths, spanning a large portion of the photosphere, including the regions of formation of molecular lines typically used in abundance work ($\log\tau_\mathrm{\,Ross}\simeq-2$).

Both our 3D model and line synthesis assumed LTE, yet the impact of departures from the LTE approximation is likely to be very important at low metallicities \citep[e.g.,][]{asplund05:review}. To first order, however, non-LTE affects the abundances but not the shapes of line profiles. A fully consistent 3D non-LTE model atmosphere and line formation scheme is currently beyond our reach. Nevertheless, our work represents one step forward in the direction of major improvements to elemental abundance determination techniques.

\acknowledgments

This work was supported in part by the Robert A. Welch Foundation of Houston, Texas (grant number F-634). We thank the anonymous referee for helping us improve the original manuscript.

{\it Facilities:} \facility{McD:2.7m}.

\end{document}